\documentstyle[aps,epsfig,twocolumn]{revtex}

\title{The Physical Implementation of Quantum Computation}

\author{David P. DiVincenzo}

\address{\vspace*{1.2ex} \hspace*{0.5ex}{IBM T.J. Watson Research
Center, Yorktown Heights, NY 10598 USA}}

\date{\today}

\begin{document}

\newcommand{\ket}[1]{| #1 \rangle}
\newcommand{\bra}[1]{\langle #1 |}

\twocolumn[\hsize\textwidth\columnwidth\hsize\csname @twocolumnfalse\endcsname

\maketitle

\begin{abstract}
After a brief introduction to the principles and promise of quantum
information processing, the requirements for the physical
implementation of quantum computation are discussed.  These five
requirements, plus two relating to the communication of quantum
information, are extensively explored and related to the many schemes
in atomic physics, quantum optics, nuclear and electron magnetic
resonance spectroscopy, superconducting electronics, and quantum-dot
physics, for achieving quantum computing.
\end{abstract}
\medskip
\medskip

\vskip1pc]

\narrowtext

\section{Introduction}

\footnote{Prepared for Fortschritte der Physik special issue, {\em
Experimental Proposals for Quantum Computation}, eds. H.-K. Lo and
S. Braunstein.}  The advent of quantum information processing, as an
abstract concept, has given birth to a great deal of new thinking, of
a very concrete form, about how to create physical computing devices
that operate in the hitherto unexplored quantum mechanical regime.
The efforts now underway to produce working laboratory devices that
perform this profoundly new form of information processing are the
subject of this book.

In this chapter I provide an overview of the common objectives of the
investigations reported in the remainder of this special issue.  The
scope of the approaches, proposed and underway, to the implementation of
quantum hardware is remarkable, emerging from specialties in atomic
physics\cite{atom}, in quantum optics\cite{opt}, in nuclear\cite{NMR}
and electron\cite{Kaneetc} magnetic resonance spectroscopy, in
superconducting device physics\cite{super}, in electron
physics\cite{platz}, and in mesoscopic and quantum dot
research\cite{dot}.  This amazing variety of approaches has arisen
because, as we will see, the principles of quantum computing are posed
using the most fundamental ideas of quantum mechanics, ones whose
embodiment can be contemplated in virtually every branch of quantum
physics.

The interdisciplinary spirit which has been fostered as a result is
one of the most pleasant and remarkable features of this field.  The
excitement and freshness that has been produced bodes well for the
prospect for discovery, invention, and innovation in this endeavor.

\section{Why {\it quantum\ } information processing?}

The shortest of answers to this question would be, why not?  The
manipulation and transmission of information is today carried out by
physical machines (computers, routers, scanners, etc.), in which the
embodiment and transformations of this information can be described
using the language of classical mechanics.  But the final physical
theory of the world is not Newtonian mechanics, and there is no reason
to suppose that machines following the laws of quantum mechanics
should have the same computational power as classical machines;
indeed, since Newtonian mechanics emerges as a special limit of
quantum mechanics, quantum machines can only have greater
computational power than classical ones.  The great pioneers and
visionaries who pointed the way towards quantum computers,
Deutsch\cite{Deutsch}, Feynman\cite{Feynman}, and others, were
stimulated by such thoughts.  Of course, by a similar line of
reasoning, it may well be asked whether machines embodying the
principles of other refined descriptions of nature (perhaps general
relativity or string theory) may have even more information processing
capabilities; speculations exist about these more exotic
possibilities, but they are beyond the scope of the present
discussion.

But computing with quantum mechanics really deserves a lot more
attention than wormhole computing or quantum-gravity computing; quantum
computing, while far in the future from the perspective of CMOS roadmaps
and projections of chip fab advances, can certainly be seen as a real
prospect from the perspective of research studies in quantum physics. 
It does not require science fiction to envision a quantum computer; the
proposals discussed later in this issue paint a rather definite
picture of what a real quantum computer will look like.

So, how much is gained by computing with quantum physics over computing
with classical physics?  We do not seem to be near to a final answer to
this question, which is natural since even the ultimate computing power
of classical machines remains unknown.  But the answer as we know it
today has an unexpected structure; it is not that quantum tools simply
speed up all information processing tasks by a uniform amount.  By a
standard complexity measure (i.e., the way in which the number of
computational steps required to complete a task grows with the ``size''
$n$ of the task), some tasks are not sped up at all\cite{polys} by using
quantum tools (e.g., obtaining the $n$th iterate of a function
$f(f(...f(x)...))$\cite{Ozhi}), some are sped up moderately (locating
an entry in a database of $n$ entries\cite{Gro}), and some are
apparently sped up exponentially (Shor's algorithm for factoring an
$n$-digit number\cite{PS}).

In other types of information processing tasks, particularly those
involving communication\cite{mmm200}, both quantitative and
qualitative improvements are seen\cite{firstcomm}: for certain tasks
(choosing a free day for an appointment between two parties from out
of $n$ days) there is a quadratic reduction of the amount of
communicated data required, if quantum states rather than classical
states are transmitted\cite{cleve98}.  For some tasks (the ``set
disjointness problem'', related to allocating non-overlapping segments
of a shared memory in a distributed computation) the reduction of
required communication is exponential\cite{Ambai}.  Finally, there are
tasks that are doable in the quantum world that have no counterpart
classically: quantum cryptography provides an absolute secrecy of
communication between parties that is impossible
classically\cite{BB84}.  And for some games, winning strategies become
possible with the use of quantum resources that are not available
otherwise\cite{games,gmn}.

This issue, and this chapter, are primarily concerned with the ``hows''
of quantum computing rather than the ``whys,'' so we will leave behind
the computer science after this extremely brief mention.  There is no
shortage of other places to obtain more information about these things;
I recommend the recent articles by Aharonov\cite{darev} and by
Cleve\cite{rcrev}; other general introductions\cite{revs} will give the
reader pointers to the already vast specialized literature on this
subject.

\section{Realizing quantum computation}

Let me proceed with the main topic: the physical realization of
quantum information processing.  As a guide to the remainder of the
special issue, and as a means of reviewing the basic steps required to
make quantum computation work, I can think of no better plan than to
review a set of basic criteria that my coworkers and I have been
discussing over the last few years\cite{5points} for the realization of
quantum computation (and communication), and to discuss the application
of these criteria to the multitude of physical implementations that are
found below.

So, without further ado, here are the

\bigskip

{\bf Five (plus two) requirements for the implementation of 
quantum computation}

\bigskip

{\em 1. A scalable physical system with well characterized qubits}

For a start, a physical system containing a collection of qubits is
needed.  A qubit (or, more precisely, the embodiment of a qubit) 
is\cite{Schum} simply a quantum two-level system like the two spin
states of a spin 1/2 particle, like the ground and excited states of an
atom, or like the vertical and horizontal polarization of a single
photon.  The generic notation for a qubit state denotes one state as
$\ket{0}$ and the other as $\ket{1}$.  The essential feature that
distinguishes a qubit from a bit is that, according to the laws of
quantum mechanics, the permitted states of a single qubit fills up a
two-dimensional complex vector space; the general state is written
$a\ket{0}+b\ket{1}$, where $a$ and $b$ are complex numbers, and a
normalization convention $|a|^2+|b|^2=1$ is normally adopted.  The
general state of two qubits, $a\ket{00}+b\ket{01}+c\ket{10}+d\ket{11}$,
is a four-dimensional vector, one dimension for each distinguishable
state of the two systems.  These states are generically {\em entangled},
meaning that they cannot be written as a product of the states of
two individual qubits.  The general state of $n$ qubits is specified by
a $2^n$-dimensional complex vector.

A qubit being ``well characterized'' means several different things.
Its physical parameters should be accurately known, including the
internal Hamiltonian of the qubit (which determines the energy
eigenstates of the qubit, which are often, although not always, taken
as the $\ket{0}$ and $\ket{1}$ states), the presence of and couplings
to other states of the qubit, the interactions with other qubits, and
the couplings to external fields that might be used to manipulate the
state of the qubit.  If the qubit has third, fourth, etc., levels, the
computer's control apparatus should be designed so that the
probability of the system ever going into these states is small.  The
smallness of this and other parameters will be determined by the
capabilities of quantum error correction, which will be discussed
under requirement 3.

Recognizing a qubit can be trickier than one might think.  For
example, we might consider a pair of one-electron quantum dots that
share a single electron between them as a two-qubit system.  It is
certainly true that we can denote the presence or absence of an
electron on each dot by $\ket{0}$ and $\ket{1}$, and it is well known
experimentally how to put this system into the ``entangled'' state
$1/\sqrt{2}(\ket{01}+\ket{10})$ in which the electron is in a
superposition of being on the left dot and the right dot.  But it is
fallacious to consider this as a two-qubit system; while the states
$\ket{00}$ and $\ket{11}$ are other allowed physical states of the
dots, superselection principles forbid the creation of entangled
states involving different particle numbers such as
$1/\sqrt{2}(\ket{00}+\ket{11})$.  

It is therefore false to consider this as a two-qubit system, and,
since there are not two qubits, it is nonsense to say that there is
entanglement in this system.  It is correct to say that the electron
is in a superposition of different quantum states living on the two
different dots.  It is also perfectly correct to consider this system
to be the embodiment of a {\em single} qubit, spanned by the states
(in the misleading notation above) $\ket{01}$ (``electron on the right
dot'') and $\ket{10}$ (``electron on the left dot'').  Indeed, several
of the viable proposals, including the ones by Sch{\"o}n, Averin, and
Tanamoto in this special issue, use exactly this system as a qubit.
However, false lines of reasoning like the one outlined here have sunk
various proposals before they were properly launched (no such abortive
proposals are represented in this book, but they can be found
occasionally in the literature).

An amazing variety of realizations of the qubit are represented in
this volume.  There is a very well developed line of work that began
with the proposal of Cirac and Zoller\cite{atom} for an ion-trap
quantum computer, in which, in its quiescent state, the computer holds
the qubits in pairs of energy levels of ions held in a linear
electromagnetic trap.  Various pairs of energy levels (e.g.,
Zeeman-degenerate ground states, as are also used in the NMR
approach\cite{NMR} discussed by Cory) have been proposed and
investigated experimentally. The many neutral-atom proposals (see
chapters by Kimble\cite{opt}, Deutsch\cite{lat1}, and
Briegel\cite{lat2}) use similar atomic energy levels of neutral
species.  These atomic-physics based proposals use other auxiliary
qubits such as the position of atoms in a trap or lattice, the
presence or absence of a photon in an optical cavity, or the
vibrational quanta of trapped electrons, ions or atoms (in the
Platzman proposal below\cite{platz} this is the primary qubit).  Many
of the solid-state proposals exploit the fact that impurities or
quantum dots have well characterized discrete energy level spectra;
these include the spin states of quantum dots (see chapters by
Loss\cite{dot} and Imamoglu\cite{opt}), the spin states of donor
impurities (see Kane\cite{Kaneetc}), and the orbital or charge states
of quantum dots (see Tanamoto\cite{dot}). Finally, there are a variety
of interesting proposals which use the quantized states of
superconducting devices, either ones involving the (Cooper-pair)
charge (see Sch{\"o}n, Averin), or the flux (see Mooij)\cite{super}.

\medskip

{\em 2. The ability to initialize the state of the qubits to a simple
fiducial state, such as $\ket{000...}$}

This arises first from the straightforward computing requirement that
registers should be initialized to a known value before the start of
computation.  There is a second reason for this initialization
requirement: quantum error correction (see requirement 3 below)
requires a continuous, fresh supply of qubits in a low-entropy state
(like the $\ket{0}$ state).  The need for a continuous supply of 0s,
rather than just an initial supply, is a real headache for many
proposed implementations.  But since it is likely that a demonstration
of a substantial degree of quantum error correction is still quite
some time off, the problem of continuous initialization does not have
to be solved very soon; still, experimentalists should be aware that
the speed with which a qubit can be zeroed will eventually be a very
important issue.  If the time it takes to do this initialization is
relatively long compared with gate-operation times (see requirement
4), then the quantum computer will have to equipped with some kind of
``qubit conveyor belt'', on which qubits in need of initialization are
carried away from the region in which active computation is taking
place, initialized while on the ``belt'', then brought back to the
active place after the initialization is finished.  (A similar parade
of qubits will be envisioned in requirement 5 for the case of low
quantum-efficiency measurements\cite{grate}.)

There are two main approaches to setting qubits to a standard state:
the system can either be ``naturally'' cooled when the ground state of
its Hamiltonian is the state of interest, or the standard state can be
achieved by a measurement which projects the system either into the
state desired or another state which can be rotated into it.  These
approaches are not fundamentally different from one another, since the
projection procedure is a form of cooling; for instance, the laser
cooling techniques used routinely now for the cooling of ion states to
near their ground state in a trap\cite{atom} are closely connected to
the fluorescence techniques used to measure the state of these ions.
A more ``natural'' kind of cooling is advocated in many of the
electron spin resonance based techniques (using quantum dots or
impurities)\cite{dot,Kaneetc} in which the spins are placed in a
strong magnetic field and allowed to align with it via interaction
with their heat bath.  In this kind of approach the time scale will be
a problem.  Since the natural thermalization times are never shorter
than the decoherence time of the system, this procedure will be too
slow for the needs of error correction and a ``conveyor belt'' scheme
would be required.  Cooling by projection, in which the Hamiltonian of
the system and its environment are necessarily perturbed strongly,
will have a time scale dependent on the details of the setup, but
potentially much shorter than the natural relaxation times.  One
cannot say too much more at this point, as the schemes for measurement
have in many cases not been fully implemented (see requirement 5).  In
the NMR quantum computer implementations to date (see Cory below),
cooling of the initial state has been foregone altogether; it is
acknowledged\cite{NMR} that until some of the proposed cooling schemes
are implemented (a nontrivial thing to do), NMR can never be a
scalable scheme for quantum computing.

\medskip

{\em 3. Long relevant decoherence times, much longer than the gate
operation time}

Decoherence times characterize the dynamics of a qubit (or any quantum
system) in contact with its environment.  The (somewhat overly)
simplified definition of this time is that it is the characteristic
time for a generic qubit state $\ket{\psi}=a\ket{0}+b\ket{1}$ to be
transformed into the mixture
$\rho=|a|^2\ket{0}\bra{0}+|b|^2\ket{1}\bra{1}$.  A more proper
characterization of decoherence, in which the decay can depend on the
form of the initial state, in which the state amplitudes may change as
well, and in which other quantum states of the qubit can play a role
(in a special form of state decay called ``leakage'' in quantum
computing\cite{PK1,PK2}), is rather more technical than I want to get
here; but see Refs. \cite{presk} and \cite{Chua} for a good general
discussion of all these.  Even the simplest discussion of decoherence
that I have given here should also be extended to include the
possibility that the decoherence of neighboring qubits is correlated.
It seems safest to assume that they will be neither completely
correlated nor completely uncorrelated, and the thinking about error
correction has taken this into account.

Decoherence is very important for the fundamentals of quantum physics,
as it is identified as the principal mechanism for the emergence of
classical behavior.  For the same reason, decoherence is very
dangerous for quantum computing, since if it acts for very long, the
capability of the quantum computer will not be so different from that
of a classical machine.  The decoherence time must be long enough that
the uniquely quantum features of this style of computation have a
chance to come into play.  How long is ``long enough'' is also
indicated by the results of quantum error correction, which I will
summarize shortly.

I have indicated that the ``relevant'' decoherence times should be
long enough.  This emphasizes that a quantum particle can have many
decoherence times pertaining to different degrees of freedom of that
particle.  But many of these can be irrelevant to the functioning of
this particle as a qubit.  For example, the rapid decoherence of an
electron's position state in a solid state environment does not
preclude its having a very long spin coherence time, and it can be
arranged that this is the only time relevant for quantum computation.
Which time is relevant is determined by the choice of the qubit basis
states $\ket{0}$ and $\ket{1}$; for example, if these two states
correspond to different spin states but identical orbital states, then
orbital decoherence will be irrelevant.

One might worry that the decoherence time necessary to do a successful
quantum computation will scale with the duration of the computation.
This would place incredibly stringent requirements on the physical
system implementing the computation.  Fortunately, in one of the great
discoveries of quantum information theory (in 1995-6), it was found
that error correction of quantum states is possible\cite{ShorSteane}
and that this correction procedure can be successfully applied in
quantum computation\cite{FT}, putting much more reasonable (although
still daunting) requirements on the needed decoherence times.

In brief, quantum error correction starts with coding; as in binary
error correction codes, in which only a subset of all boolean strings
are ``legal'' states, quantum error correction codes consist of legal
states confined to a subspace of the vector space of a collection of
qubits.  Departure from this subspace is caused by decoherence.  Codes
can be chosen such that, with a suitable sequence of quantum
computations and measurements of some ancillary qubits, the error caused
by decoherence can be detected and corrected.  As noted above, these
ancillary qubits have to be continuously refreshed for use.  I will not
go much farther into the subject here, see \cite{presk} for more.  It is
known that quantum error correction can be made fully fault tolerant,
meaning that error correction operations can be successfully
intermingled with quantum computation operations, that errors occurring
during the act of error correction, if they occur at a sufficiently
small rate, do no harm, and that the act of quantum computation does not
itself cause an unreasonable proliferation of errors.

These detailed analyses have indicated the magnitude of decoherence time
scales that are acceptable for fault-tolerant quantum computation.  The
result is that, if the decoherence time is $10^4-10^5$ times the ``clock
time'' of the quantum computer, that is, the time for the execution of
an individual quantum gate (see requirement 4), then error correction
can be successful.  This is, to tell the truth, a rather stringent
condition, quantum systems frequently do not have such long decoherence
times.  But sometimes they do, and our search for a successful physical
implementation must turn towards these.  At least this result says that
the required decoherence rate does not become ever smaller as the size
and duration of the quantum computation grows.  So, once the desired
threshold is attainable, decoherence will not be an obstacle to scalable
quantum computation.

Having said this, it must be admitted that it will be some time before
it is even possible to subject quantum error correction to a reasonable
test.  Nearly all parts of requirements 1-5 must be in place before such
a test is possible.  And even the most limited application of quantum
error correction has quite a large overhead: roughly 10 ancillary qubits
must be added for each individual qubit of the computation. 
Fortunately, this overhead ratio grows only logarithmically as the the
size of the quantum computation is increased.

In the short run, it is at least possible to design and perform
experiments which measure the decoherence times and other relevant
properties (such as the correlation of decoherence of neighboring
qubits) of candidate implementations of qubits.  With such initial
test experiments, caution must be exercised in interpreting the
results, because decoherence is a very system-specific phenomenon,
depending on the details of all the qubits' couplings to various
environmental degrees of freedom.  For example, the decoherence time
of the spin of an impurity in the bulk of a perfect semiconductor may
not be the same as its decoherence time when it is near the surface of
the solid, in the immediate neighborhood of device structures designed
to manipulate its quantum state.  Test experiments should probe
decoherence in as realistic a structure as is possible.

\medskip

{\em 4. A ``universal'' set of quantum gates}

This requirement is of course at the heart of quantum computing.  A
quantum algorithm is typically specified\cite{Deutsch} as a sequence
of unitary transformations $U_1$, $U_2$, $U_3$,..., each acting on a
small number of qubits, typically no more than three.  The most
straightforward transcription of this into a physical specification is
to identify Hamiltonians which generate these unitary transformations,
viz., $U_1=e^{iH_1t/\hbar}$, $U_2=e^{iH_2t/\hbar}$,
$U_3=e^{iH_3t/\hbar}$, etc.; then, the physical apparatus should be
designed so that $H_1$ can be turned on from time $0$ to time $t$,
then turned off and $H_2$ turned on from time $t$ to time $2t$, etc.

Would that life were so simple!  In reality what can be done is much
less, but much less can be sufficient.  Understanding exactly how
much less is still enough, is the main complication of this requirement.
In all the physical implementations discussed in this
volume, only particular sorts of Hamiltonians can be turned on
and off; in most cases, for example, only two-body (two-qubit)
interactions are considered.  This immediately poses a problem for a
quantum computation specified with three-qubit unitary transformations;
fortunately, of course, these can always be re-expressed in terms of
sequences of one- and two-body interactions\cite{2bit}, and the two-body
interactions can be of just one type\cite{Bar}, the ``quantum XOR'' or
``cNOT''.  There are some implementations in which multi-qubit gates
can be implemented directly\cite{Mol2}.

However, this still leaves a lot of work to do.  In some systems,
notably in NMR (see Cory), there are two-body interactions present which
cannot be turned {\em off}, as well as others which are switchable. 
This would in general be fatal for quantum computation, but the
particular form of the fixed interactions permit their  effects to be
annulled by particular ``refocusing'' sequences of the controllable
interactions, and it has recently been discovered\cite{Debbie} that
these refocusing sequences can be designed and implemented efficiently.

For many other systems, the two-body Hamiltonian needed to generate
directly the cNOT unitary transformation is not available.  For example,
in the quantum-dot proposal described by Loss below\cite{dot}, the only
two-body interaction which should be easily achievable is the exchange
interaction between neighboring spins, $H\propto{\vec S}_i\cdot{\vec
S}_{i+1}$; in the Imamoglu chapter\cite{opt}, the attainable
interaction is of the XY type, i.e., $H\propto
S_{ix}S_{jx}+S_{iy}S_{jy}$.  An important observation is that with the
appropriate sequence of exchange or XY interactions, in conjunction with
particular one-body interactions (which are assumed to be more easily
doable), the cNOT transformation can be synthesized\cite{Burk}.  It is
incumbent on each implementation proposal to exhibit such a sequence for
producing the cNOT using the interactions that are naturally realizable.

Often there is also some sophisticated thinking required about the
time profile of the two-qubit interaction.  The naive description
above uses a ``square pulse'' time profile, but often this is
completely inappropriate; for instance, if the Hamiltonian can also
couple the qubit to other, higher-lying levels of the quantum system,
often the only way to get the desired transformation is to turn on and
off the interaction smoothly and slowly enough that an adiabatic
approximation is accurate\cite{PK1,PK2} (in a solid-state context, see
also \cite{Burk2}).  The actual duration of the pulse will have to be
sufficiently long that any such adiabatic requirement is satisfied;
then typically only the time integral $\int dt H(t)$ is relevant for
the quantum gate action.  The overall time scale of the interaction
pulse is also controlled by the attainable maximum size of the matrix
elements of $H(t)$, which will be determined by various fundamental
considerations, like the requirement that the system remains in the
regime of validity of a linear approximation, and practical
considerations, like the laser power that can be concentrated on a
particular ion.  Given these various constraints, the ``clock time''
of the quantum computer will be determined by the time interval needed
such that two consecutive pulses have negligible overlap.

Another consideration, which does not seem to present a problem with
any current implementation schemes, but which may be an issue in the
future, is the classicality of the control apparatus.  We say that the
interaction Hamiltonian $H(t)$ has a time profile which is controlled
externally by some ``classical'' means, that is, by the intensity of a
laser beam, the value of a gate voltage, or the current level in a wire.
 But each of these control devices is made up themselves of quantum
mechanical parts.  When we require that these behave classically, it
means that their action should proceed without any entanglement
developing between these control devices and the quantum computer.
Estimates indicate that this entanglement can indeed be negligible, but
this effect needs to be assessed for each individual case.

In many cases it is impossible to turn on the desired interaction
between a pair of qubits; for instance, in the ion-trap scheme, no
direct interaction is available between the ion-level qubits\cite{atom}. 
In this and in other cases, a special quantum subsystem (sometimes
referred to as a ``bus qubit'') is used which can interact with each of
the qubits in turn and mediate the desired interaction: for the
ion trap, this is envisioned to be the vibrational state of the
ion chain in the trap; in other cases it is a cavity photon whose
wavefunction overlaps all the qubits.  Unfortunately, this auxiliary
quantum system introduces new channels for the environment to couple to
the system and cause decoherence, and indeed the decoherence occurring
during gate operation is of concern in the ion-trap and
cavity-quantum electrodynamics schemes.

Some points about requirement 4 are important to note in relation to the
implementation of error correction.  Successful error correction
requires fully parallel operation, meaning that gate operations
involving a finite fraction of all the qubits must be doable
simultaneously.  This can present a problem with some of the proposals in
which the single ``bus qubit'' is needed to mediate each interaction.
On the other hand, the constraint that interactions are only among
nearest neighbors in a lattice, as in many of the solid-state
proposals, does allow for sufficient parallelism\cite{Gott}.

Quantum gates cannot be implemented perfectly; we must expect both
systematic and random errors in the implementation of the necessary
Hamiltonians.  Both types of errors can be viewed as another source of
decoherence and thus error correction techniques are effective for
producing reliable computations from unreliable gates, if the
unreliability is small enough.  The tolerable unreliability due to
random errors is in the same vicinity as the decoherence threshold, that
is, the magnitude of random errors should be $10^{-4}-10^{-5}$ per gate
operation or so.  It might be hoped that systematic errors could be
virtually eliminated by careful calibration; but this will surely not
always be the case.  It seems harder to give a good rule for how much
systematic error is tolerable, the conservative estimates give a very,
very small number (the square of the above)\cite{presk}, but on the
other hand there seems to be some evidence that certain important
quantum computations (e.g., the quantum Fourier transform) can tolerate
a very high level of systematic error (over- or under-rotation).  Some
types of very large errors may be tolerable if their presence can be
detected and accounted for on the fly (we are thinking, for example,
about charge switching in semiconductors or superconductors).

Error correction requires that gate operations be done on coded qubits,
and one might worry that such operations would require a new repertoire
of elementary gate operations for the base-level qubits which make up
the code.  For the most important error correction techniques, using the
so called ``stabilizer'' codes, this is not the case.  The base-level
toolkit is exactly the same as for the unencoded case: one-bit gates and
cNOTs, or any gate repertoire that can produce these, are adequate. 
Sometimes the use of coding can actually {\em reduce} the gate
repertoire required: in the work on decoherence free subspaces and
subsystems, codes are introduced using blocks of three and four qubits
for which two-qubit exchange interactions alone are enough to implement
general quantum computation\cite{Bac,latest}.  This simplification could
be very useful in the quantum-dot\cite{dot} or
semiconductor-impurity\cite{Kaneetc} implementations.

\medskip

{\em 5. A qubit-specific measurement capability}

Finally, the result of a computation must be read out, and this requires
the ability to measure specific qubits.  In an ideal measurement, if a
qubit's density matrix is
$\rho=p\ket{0}\bra{0}+(1-p)\ket{1}\bra{1}+\alpha\ket{0}\bra{1}+
\alpha^*\ket{1}\bra{0}$, the measurement should give outcome ``0'' with
probability $p$ and ``1'' with probability $1-p$ independent of $\alpha$
and of any other parameters of the system, including the state of nearby
qubits, and without changing the state of the rest of the quantum
computer.  If the measurement is ``non-demolition'', that is, if in
addition to reporting outcome ``0'' the measurement leaves the qubit in
state $\ket{0}\bra{0}$, then it can also be used for the state
preparation of requirement 2; but requirement 2 can be  fulfilled in
other ways.  

Such an ideal measurement as I have described is said to have 100\%
quantum efficiency; real measurements always have less.  While the
fidelity of a quantum measurement is not captured by a single number,
the single quantum-efficiency parameter is often a very useful way to
summarize it, just as the decoherence time is a useful if incomplete
summary of the damage caused to a quantum state by the environment.

While quantum efficiency of 100\% is desirable, much less is needed
for quantum computation; there is, in fact, a tradeoff possible
between quantum efficiency and other resources which results in
reliable computation.  As a simple example, if the quantum efficiency
is 90\%, then, in the absence of any other imperfections, a
computation with a single-bit output (a so-called ``decision
problem'', common in computer science) will have 90\% reliability. If
97\% reliability is needed, this can just be achieved by rerunning the
calculation three times.  Much better, actually, is to ``copy'' the
single output qubit to three, by applying two cNOT gates involving the
output qubit and two other qubits set to $\ket{0}$, and measuring
those three.  (Of course, qubits cannot be ``copied'', but their value
in a particular basis can.)  In general, if quantum efficiency $q$ is
available, then copying to somewhat more than $1/q$ qubits and
measuring all of these will result in a reliable outcome.  So, a
quantum efficiency of 1\% would be usable for quantum computation, at
the expense of hundreds of copies/remeasures of the same output qubit.
(This assumes that the measurement does not otherwise disturb the
quantum computer.  If it does, the possibilities are much more
limited.)

Even quantum efficiencies much, much lower than 1\% can be and are
used for successful quantum computation: this is the ``bulk'' model of
NMR (see Cory and \cite{NMR}), where macroscopic numbers of copies of
the same quantum computer (different molecules in solution) run
simultaneously, with the final measurement done as an ensemble average
over the whole sample.  These kinds of weak measurements, in which
each individual qubit is hardly disturbed, are quite common and well
understood in condensed-matter physics.

If a measurement can be completed quickly, on the timescale of
$10^{-4}$ of the decoherence time, say, then its repeated application
during the course of quantum computation is valuable for simplifying
the process of quantum error correction.  On the other hand, if this
fast measurement capability is not available, quantum error correction
is still possible, but it then requires a greater number of quantum
gates to implement.

Other tradeoffs between the complexity and reliability of quantum
measurement vs. those of quantum computation have recently been explored.
 It has been shown that if qubits can be initialized into pairs of
maximally entangled states, and two-qubit measurements in the so-called
Bell basis ($\Psi^\pm=\ket{01}\pm\ket{10}$,
$\Phi^\pm=\ket{00}\pm\ket{11}$) are possible, then no two-qubit quantum
gates are needed, one-bit gates alone suffice\cite{Nat}.  Now, often
this tradeoff will not be useful, as in many schemes a Bell measurement
would require two-bit quantum gates.

But the overall message, seen in many of our requirements, is that more
and more, the theoretical study of quantum computation has offered a
great variety of tradeoffs for the potential implementations: if X is
very hard, it can be substituted with more of Y.  Of course, in many
cases both X and Y are beyond the present experimental state of the
art; but a thorough knowledge of these tradeoffs should be very useful
for devising a rational plan for the pursuit of future experiments.

\section{Desiderata for quantum communication}

For computation alone, the five requirements above suffice.  But the
advantages of quantum information processing are not manifest solely, or
perhaps even principally, for straightforward computation only.  There
are many kinds of information-processing tasks, reviewed briefly at the
beginning, that involve more than just computation, and for which
quantum tools provide a unique advantage.

The tasks we have in mind here all involve not only computation but
also communication.  The list of these tasks that have been
considered in the light of quantum capabilities, and for which some
advantage has been found in using quantum tools, is fairly long and
diverse: it includes secret key distribution, multiparty function
evaluation as in appointment scheduling, secret sharing, and game
playing\cite{mmm200}.

When we say communication we mean quantum communication: the transmission
of intact qubits from place to place.  This obviously adds more
features that the physical apparatus must have to carry out this
information processing.  We formalize these by adding two more items to
the list of requirements:

\medskip

{\em 6. The ability to interconvert stationary and flying qubits}

\medskip

{\em 7. The ability faithfully to transmit flying qubits between
specified locations}

These two requirements are obviously closely related, but it is
worthwhile to consider them separately, because some tasks need one but
not the other.  For instance, quantum cryptography\cite{BB84} involves
only requirement 7; it is sufficient to create and detect flying qubits
directly.

I have used the jargon ``flying qubits''\cite{opt}, which has become
current in the discussions of quantum communication.  Using this term
emphasizes that the optimal embodiment of qubits that are readily
transmitted from place to place is likely to be very different from the
optimal qubits for reliable local computation.  Indeed, almost all
proposals assume that photon states, with the qubit encoded either in
the polarization or in the spatial wavefunction of the photon, will be
the flying qubit of choice, and indeed, the well developed technology of
light transmission through optical fibers provides a very promising
system for the transmission of qubits.  I would note, though, that my
colleagues and I have raised the possibility that electrons traveling
though solids could provide another realization of the flying 
qubit\cite{mmm200,NewLoss}.

Only a few completely developed proposals exist which incorporate
requirements 6 and 7.  Of course, there are a number of quite detailed
studies of 7, in the sense that experiments on quantum cryptography have
been very concerned with the preservation of the photon quantum state
during transmission through optical fibers or through the atmosphere. 
However, these studies are rather disconnected from the other concerns
of quantum computing.  Requirement 6 is the really hard one; to date the
only theoretical proposal sufficiently concrete that experiments
addressing it have been planned is the scheme produced by Kimble and
coworkers\cite{photon} for unloading a cavity photon into a traveling
mode via atomic spectroscopy, and loading it by the time-reversed
process.  Other promising concepts, like the launching of electrons from
quantum dots into quantum wires such that the spin coherence of the
electrons is preserved, need to be worked out more fully.

\section{Summary}

So, what is the ``winning'' technology going to be?  I don't think that
any living mortal has an answer to this question, and at this point
it may be counterproductive even to ask it.  Even though we have
lived with quantum mechanics for a century, our study of quantum effects
in complex artificial systems like those we have in mind for quantum
computing is in its infancy.  No one can see how or whether all the
requirements above can be fulfilled, or whether there are new tradeoffs,
not envisioned in our present theoretical discussions but suggested by
further experiments, that might take our investigations
in an entirely new path.

Indeed, the above discussion, and the other chapters of this special
issue, really do not cover all the foreseeable approaches.  I will
mention two of which I am aware: first, another computational
paradigm, that of the cellular automaton, is potentially available for
exploitation.  This is distinguished from the above ``general
purpose'' approach in that it assumes that every bit pattern
throughout the computer will be subjected to the same evolution rule.
It is known that general-purpose computation is performable, although
with considerable overhead, by a cellular automaton.  This is true as
well for the quantum version of the cellular automaton, as
Lloyd\cite{L93} indicated in his original work.  New theoretical work
by Benjamin\cite{Benj} shows very explicitly how relatively simple
local rules would permit the implementation of some quantum
computations.  This could point us perhaps towards some sort of
polymer with a string of qubits on its backbone that can be addressed
globally in a spectroscopic fashion.  Experiments are not
oriented towards this at the moment, but the tradeoffs are very
different, and I don't believe it should be excluded in the future.

Second, even more speculative, but very elegant, is the proposal of
Kitaev\cite{Kitaev} to use quantum systems with particular kinds of
topological excitations, for example nonabelian anyons, for quantum
computing.  It is hard to see at the moment how to turn this exciting
proposal into an experimental program, as no known physical system is
agreed to have the appropriate topological excitations.  But further
research in, for example, the quantum Hall effect might reveal such a
system; more likely, perhaps, is that further understanding of this
approach, and that of Freedman and his colleagues\cite{Fried}, will shed
more light on doing quantum computing using the ``standard'' approach
being considered in this book.

I am convinced of one thing: the ideas of quantum information theory
will continue to exert a decisive influence on the further
investigation of the fundamental quantum properties of complex quantum
systems, and will stimulate many creative and exciting developments
for many years to come.

\section*{Acknowledgments}

I gratefully acknowledge support from the Army Research Office under
contract number DAAG55-98-C-0041.  I thank Alec Maassen van den Brink
for a careful reading of this manuscript.



\end{document}